*To be submitted to Phys. Rev. Lett.*

**Probing the role of single defects on the thermodynamics of electric-field induced phase transitions**

S.V. Kalinin,[1,2,*] S. Jesse,[1] B.J. Rodriguez,[1,2] Y.H. Chu,[3] R. Ramesh,[3] E.A. Eliseev[4] and A.N. Morozovska[4]

[1]Materials Sciences and Technology Division and [2]The Center for Nanophase Materials Sciences, Oak Ridge National Laboratory, Oak Ridge, TN 37831

[3]Department of Physics and Department of Materials Science and Engineering, University of California, Berkeley, CA 94720

[4]National Academy of Science of Ukraine, Kiev, Ukraine

The kinetics and thermodynamics of first order transitions is universally controlled by defects that act as nucleation sites and pinning centers. Here we demonstrate that defect-domain interactions during polarization reversal processes in ferroelectric materials result in a pronounced fine structure in electromechanical hysteresis loops. Spatially-resolved imaging of a single defect center in multiferroic $BiFeO_3$ thin film is achieved, and the defect size and built-in field are determined self-consistently from the single-point spectroscopic measurements and spatially-resolved images. This methodology is universal and can be applied to other reversible bias-induced transitions including electrochemical reactions.

[*] Corresponding author, sergei2@ornl.gov

Polarization switching in ferroelectric and multiferroic materials is the functional basis of applications including non-volatile memories[1,2] and data storage.[3] These applications have stimulated extensive effort at understanding of thermodynamic stability of ferroelectric phase in constrained geometries and fundamental mechanisms for polarization reversal. Static domain structures and polarization distributions in low-dimensional systems and in the vicinity of bulk and interfacial defects has been extensively studied theoretically using Ginzburg-Landau[4,5] and first principle theories.[6] A wealth of experimental information on polarization behavior at surfaces and in ultrathin films is becoming available with the emergence of grazing incidence X-ray methods[7,8] and electron microscopies.[9]

The experimental and theoretical progress in understanding static polarization behavior is belied by the lack of knowledge on dynamic polarization behavior. The seminal work by Landauer has demonstrated that the experimentally observed switching fields correspond to impossibly large (~$10^3$ - $10^5$ kT) values for the nucleation activation energy.[10] This discrepancy is resolved by postulating the presence of discrete switching centers initiating low-field nucleation. The spatial and energy distribution of the nucleation center has become the central component of all statistical theories for polarization switching in macroscopic ferroelectrics.[11,12] Notably, this behavior is universal to all first order phase transitions including solidification, martensitic phase transformations, and phenomena such as plastic deformation and electrochemical reactions, in which formation of a new phase is always initiated on a specific defect sites or is controlled by defect dynamics.

Recently, development of Piezoresponse Force Microscopy and focused X-ray imaging has allowed stroboscopic (PFM)[13,14] and real-time (X-ray)[15] measurements of domain growth in the uniform field of a capacitor structures and visualization of nucleation sites. On free surfaces, localized switching by PFM has been used to directly measure domain wall geometry and growth rate, directly related to the disorder in the system.[16,17] The spatially resolved *imaging* studies are complemented by *spectroscopic* studies of switching process, in which the measured electromechanical response as a function of dc tip bias provides the information on the size of the domain formed below the tip. Quantitative Piezoresponse Force Spectroscopy have demonstrated that polarization switching in the non-uniform PFM tip field on a nearly-ideal surface is close to intrinsic.[18] The variation of nucleation biases along the surface has been used to map the random field and random bond components of disorder potential, and establish the role of ferroelastic domain walls on switching.[19] However, the key question – whether the effect of a *single* localized defect on local polarization switching can be determined – has not been resolved yet.

The observation of polarization switching on a single defect level requires (a) knowledge of defect signature on local PFS spectrum, (b) sufficiently high sensitivity to detect a single defect, and (c) model system with the spacing between the defects larger then characteristic spatial resolution. These conditions are similar to those in e.g. single molecule optical microscopy, in which molecular size is well below spatial resolution, but detection of single molecules is possible due to large molecule-molecule separation. Here we predict the defect signature on local piezoresponse force spectrum, and demonstrate experimental observation of a single defect.

Simple estimates suggest that characteristic defect size responsible for polarization switching in ferroelectric is ~1-2 nm, well below PFM resolution (> 10 nm). As a model system with low density of structural defects and high surface stability, we have chosen epitaxial (100) $BiFeO_3$ thin films grown as described elsewhere.[20] Note the nearly ideal surface with roughness of 1.3 nm (after 1st order flattening) devoid of any visible topographic defects, as shown in Fig. 1 (a). The spatial variability of switching behavior is studied by Switching Spectroscopy PFM.[21] The majority of the sample surface exhibits nearly ideal hysteretic behavior (not shown). However, in several locations the recorded loops exhibit a pronounced fine structure. Several examples are shown in Fig. 1 (d), in which single, double, and even triple shoulders are evident. This behavior can be further illustrated in the first derivative of the hysteresis loops, as shown in Figs. 1 (e-g). Notably, these fine structure features are highly reproducible (up to 256 loops have been collected at a single point with negligible variation between them) and also change gradually between adjacent locations.

The observed loop fine structure suggests local deviations in the switching process from that in an ideal material due to defect-domain interactions. To determine the effect of defect on polarization switching, and establish the relationship between the defect parameters and the measured signal, we analyze the switching process within the framework of Landauer-Molotskii model.[10,22] The free energy of the semi-ellipsoidal domain is:

$$\Phi(r,l) \cong \psi_S S(r,l) - \frac{P_S}{2}\int_V dv\left(E_3^p(\mathbf{x}) + E_3^d(\mathbf{x})\right) + \frac{4\pi l r^2}{3\varepsilon_0\varepsilon_{11}} n_D(r,l) P_S^2 \ , \tag{1}$$

where $r$ is the domain radius, $l$ is its length, $S$ –surface area, $\psi_S$ - is the domain wall energy density, $P_S$ is the magnitude of spontaneous polarization. The rigorous expression for the depolarization factor $n_D(r,l)$ is given in Ref. [23]. The electric field established by the probe is $E_3^p(\mathbf{x}) = -\nabla\varphi_p(\mathbf{x})$, and the electric field created by the defects is $E_3^d(\mathbf{x}) = -\nabla\varphi_d(\mathbf{x})$ (random bond frozen disorder, FD). Note that disorder components can contribute differently to switching between different states, i.e. for 180° switching, the $\Phi_{FD}^{+}$ for $+P \to -P$ is not necessarily equal to $\Phi_{FD}^{-}$ for $-P \to +P$.

Following Tagantsev (Ref. [24]), we analyze the switching in the presence of a surface field defect. This model is chosen since (a) PFM probe electric field is maximal on surface, and hence surface defects affect nucleation stronger, and (b) defect concentration near the surface is typically much larger than in the bulk. To develop the analytical description, we adopt the Gaussian field distribution for the *i*-th defect, i.e. assume:

$$E_3^d(\mathbf{x}) = E_{Si}\exp\left(-\frac{(x-x_{0i})^2+(y-y_{0i})^2}{r_{di}^2}-\frac{z}{h_{di}}\right), \tag{2}$$

where $E_{Si}$ is the built-in field amplitude, $h_{di} << r_{di}$ is the penetration depth, $r_{di}$ is the characteristic radius, and $\mathbf{x}_{0i} = \{x_{0i}, y_{0i}, 0\}$ is the defect position. The analysis is performed assuming that the semi-ellipsoidal domain is axi-symmetric, but allowing for defect-induced shift of the domain center $\mathbf{b} = \{b_1, b_2, 0\}$ compared to the PFM tip apex location, $\mathbf{a} = \{a_1, a_2, 0\}$ [see Fig. 3 (a)]. Thus, $|\mathbf{a}-\mathbf{b}|$ is the tip-domain separation. The domain sizes *r*, *l* and domain position $\mathbf{b}$ are determined from the minimum of free energy in Eq. (1).[25]

The schematics of domain evolution for different tip-defect distances is summarized in Fig. (2). The numerical analysis of the switching process illustrates that under conditions $l >> d$, $l >> h_{di}$ and $r < d$ (where $d$ is the effective charge-surface separation or effective tip parameter) for large field strength $E_S \cong 10^8 - 10^{10}$ V/m the nucleation process is initiated at the defect location. On increasing the bias, the domain rapidly shifts towards the tip location. The coercive bias shift, $\Delta U_C = U_C^+ + U_C^-$, is

$$\Delta U_C(a_1, a_2) \approx -2h_{di}E_{Si}\exp\left(-\frac{(x_{0i}-a_1)^2+(y_{0i}-a_2)^2}{r_{di}^2}\right). \tag{3}$$

This shift is only weakly dependent on the spatial extent of the defect and is determined by local built-in potential only. Hence, the local nucleation biases can be predicted assuming that effective switching potential is a linear superposition of the tip and defect, $h_{di}E_{Si}$, potentials.[19]

The immediate consequence of the finiteness of defect size, $r_{di} < d$, is the emergence of the loop fine structure, as illustrated in Fig. 2. Specifically, for tip located at the defect site, the switching of the material within the defect occurs for smaller voltages compared to ideal surface and stops once domain fills the defect region. The switching of the remaining part of the sample occurs at same voltages as for ideal surface. This non-monotonic growth process results in the fine structure in hysteresis loop, the size of which is determined by the ratio of the signal generation volume of PFM and defect size. Similar behavior is anticipated when the tip is shifted compared to the defect center. However, in this case nucleation is delayed according to Eq. (3). Finally, for large tip-defect separations the defect volume does not contribute to the observed PFM signal.

Under the condition $\mathbf{b} = \mathbf{x}_{0i}$, the relative change of the piezoelectric response caused by the nearest $i$-th defect is

$$\Delta PR\left(r_{Si}\right) \approx \frac{16 r_{Si}}{3d_{33}^{*} + d_{15}} \left( \frac{3d_{33}^{*}}{\pi d + 8\left(r_{Si} + \left|\mathbf{a} - \mathbf{x}_{0i}\right|\right)} + \frac{d_{15}}{3\pi d + 8\left(r_{Si} + \left|\mathbf{a} - \mathbf{x}_{0i}\right|\right)} \right), \qquad (4)$$

The radius of surface domain state is $r_{Si} = r_{di}\sqrt{\ln\left(\frac{2 h_{di} P_S E_{Si}}{\psi_S + \left(4 h_{di} P_S^2 / 3\varepsilon_0 \varepsilon_{33}\right)}\right)}$, piezoelectric coefficient $d_{33}^{*} = d_{33} + \frac{d_{31}}{3}\left(1 + 4\nu\right)$, $\nu$ is the Poisson ratio, index $m$ indicates several possible fine structure elements for a given piezoresponse loop. Since $r_{Si} \sim r_{di}$, the fine structure height is a direct measure of the lateral size of the defect.

To verify the theoretical predictions, the spatial variability of hysteresis loop fine structure for the region in Fig. 1 is analyzed. The hysteresis loops have been fitted by a phenomenological fitting function representing the sum of the ideal loop and fine structure features for forward and reverse branches.[25] Shown in Fig. 4 (a) is an example of the hysteresis loop and corresponding fit. The nucleation bias map is illustrated in Fig. 4 (b). The distribution of the feature strength (total area below the fine structure feature) is illustrated in Fig. 4 (c,d) for forward and reverse branches. Note that Fig. 4 (b,c) illustrate the expected spatial signature of a single defect - the depression in the nucleation bias image and a ring encircling the defect in the fine structure image. Further confirmation of this signature is illustrated in averaged radial profiles in Fig. 4 (e,f). The feature is present only on the forward branch of the loop, in agreement with the random field character of the defect. Therefore, the spectroscopic maps in Fig. 4 (b,c) provide an image of a single random-field defect visualized both through the changes in the local nucleation bias, and evolution of the fine-structure feature in the PFS spectra.

Finally, we discuss the relationship between the defect parameters extracted from the image and the single-point spectrum analysis. The characteristic size of the single defect as seen on the PNB image Fig. 4 (c) is ~60 nm and built in-potential is ~ -2V. For these values, the expected magnitude of the fine structure feature is ~ 25-50%, consistent with experimental loop shown in Fig. 4 (a). In comparison, the analysis of a single-point hysteresis loop fine structure allows to estimate $h_d E_S$=-3.2 V, $r_d$ = 60 nm for a defect located in cell (25,20). Furthermore, the deconvolution suggests the presence of a second defect with $h_d E_S \approx +0.5$ V, $r_d \approx 30$ nm located in cell (5,25), as can be observed in the spatially resolved image [Fig. 4(b)]. Thus, the defect parameters in the single point spectra and images are similar, indicative of the self-consistency of the experimental data and fidelity of the theoretical model. Note that real-space data provides better insight into defect location, while spectroscopic data offers significantly higher resolution in energy space.

To summarize, we have predicted the signature of a single defect of the fine structure of electromechanical hysteresis loops in Piezoresponse Force Spectroscopy. Under optimal conditions of tip and surface stability, the measurements are shown to be highly reproducible and hysteresis loops with multiple (up to four) discernible features are demonstrated. The use of model system with low density of defects allowed the visualization of a single positive random field-type defect both in the nucleation bias and fine structure intensity images. The spatial variability of switching behavior and the analysis of fine structure feature provide self-consistent estimates for defect size and built-in potential. Remarkably, the very rich fine structure of reverse branch of hysteresis loops in Fig. 1 (e-g) suggests the presence of multiple negative field defects that cannot be resolved on spatial maps and necessitate the further development of 3D deconvolution algorithms for the SSPFM data. This will allow the collective effect of defects on switching dynamics to be visualized and probed in real space.

From a general perspective, scanning probe microscopy provides a natural platform for local studies of phase transitions and chemical reactions. The best known example is protein unfolding spectroscopy, in which a force applied by an AFM tip to an individual macromolecule acts as a stimulus for conformational changes and the molecule length provide readout of molecular state.[26,27] However, this example is unique – in cases such as pressure induced phase transitions (e.g. dislocation nucleation during indentation process) the process is irreversible, precluding systematic studies of the role of defects on switching. The analysis of hysteresis loop fine structure and its spatial variability to study defect mediated thermodynamics of first-order transitions can be applied to study other bias-induced phenomena, including crystallization-amorphization in phase change memories (PCM) or electrochemical reactions. This approach can be combined with other detection methods, e.g. Raman, optical, or electroluminescent detection, providing complementary information on the induced domain size and chemical changes in the system. The reversibility of many bias-induced transitions, combined with spectroscopic imaging, will allow systematic studies of the defect effect on a process as a function of probe-defect separation, environment, etc, opening the pathway for studies of defect-engineered systems and hence understanding of atomistic mechanisms of these transformations. Future integration with *in-situ* scanning transmission electron microscopy will allow atomistic insight into switching mechanisms.

Research was sponsored by the Division of Materials Sciences and Engineering and the Center for Nanophase Materials Sciences, Office of Basic Energy Sciences, U.S. Department of Energy with Oak Ridge National Laboratory, managed and operated by UT-Battelle, LLC.

## Figure captions

**Fig. 1.** (a) Surface topography and (b) PFM amplitude image of the $BiFeO_3$ surface before the SS-PFM experiment. (c) SS-PFM work of switching image. (d) Hysteresis loop shape is selected locations in (a). Hysteresis loop fine structure of (e) loop I, (f) loop II and (g) loop III. Shown are corresponding Gaussian fits. The positions and areas for each peak are given in parenthesis. The red circle illustrates the fine structure feature on forward branch studied in this work. Vertical scale in (a) is 20 nm.

**Fig. 2.** Domain growth in the presence of the defect. (a) For large tip-defect separation, the defect affects domain growth process only on late stages. Due to the finite size of the PFM signal generation volume, the defect does not contribute to the measured signal. Experimentally, this corresponds to (b) nearly ideal PFS loops. (c) For intermediate defect-tip separations, nucleation can be initiated at the defect (rather then tip) site, resulting in small change in PFM signal for low biases. The domain rapidly shifts towards the tip at higher voltages, resulting in fine structure in PFS. Corresponding nucleation bias is lowered compared to the ideal surface. This corresponds to the (d) pronounced fine structure feature on the PFS loops. (e) For the tip positioned at the center of defect, the nucleation starts at low voltages. The growth stops when the domain occupies the defect region, and resumes only when bias becomes large enough to induce domain growth on ideal surface. The size of the experimentally observed fine feature (f) is a direct measure of the spatial extent of the defect.

**Fig. 3.** (a) Schematics of domain nucleation in the vicinity of surface field defect. (b) Schematic dependence of normalized PR-response on the applied voltage $U$ in the vicinity of defect. The role of single Gaussian defect parameters $\{E_S, h_d, r_d, x_0\}$ on piezoresponse loop fine structure maximal height $\Delta PR$: (c) $E_S$=0.9 $10^9$ V/m, $h_d$=0.4, 1, 2, 3 nm (curves 1, 2, 3, 4); (d) $r_d$=10 nm, $h_d$=0.4, 1, 2, 3 nm (curves 1, 2, 3, 4); (e) $E_S$=0.9 $10^9$ V/m, $r_d$=10 nm, $h_d$=0.4, 0.6, 1.2, 3 nm (curves 1, 2, 3, 4). Material parameters correspond to $BiFeO_3$ (Ref. 25), $d$=30nm.

**Fig. 4.** (a) Schematics of the hysteresis loop with a single fine structure feature. 2D maps of (b) negative nucleation bias and integral fine structure for (c) forward and (d) reverse branch. Note the depression in (d) and associated ring structure in (c). Radial averages for (e) nucleation biases and (f) integral fine structure taken with respect to center of defect.

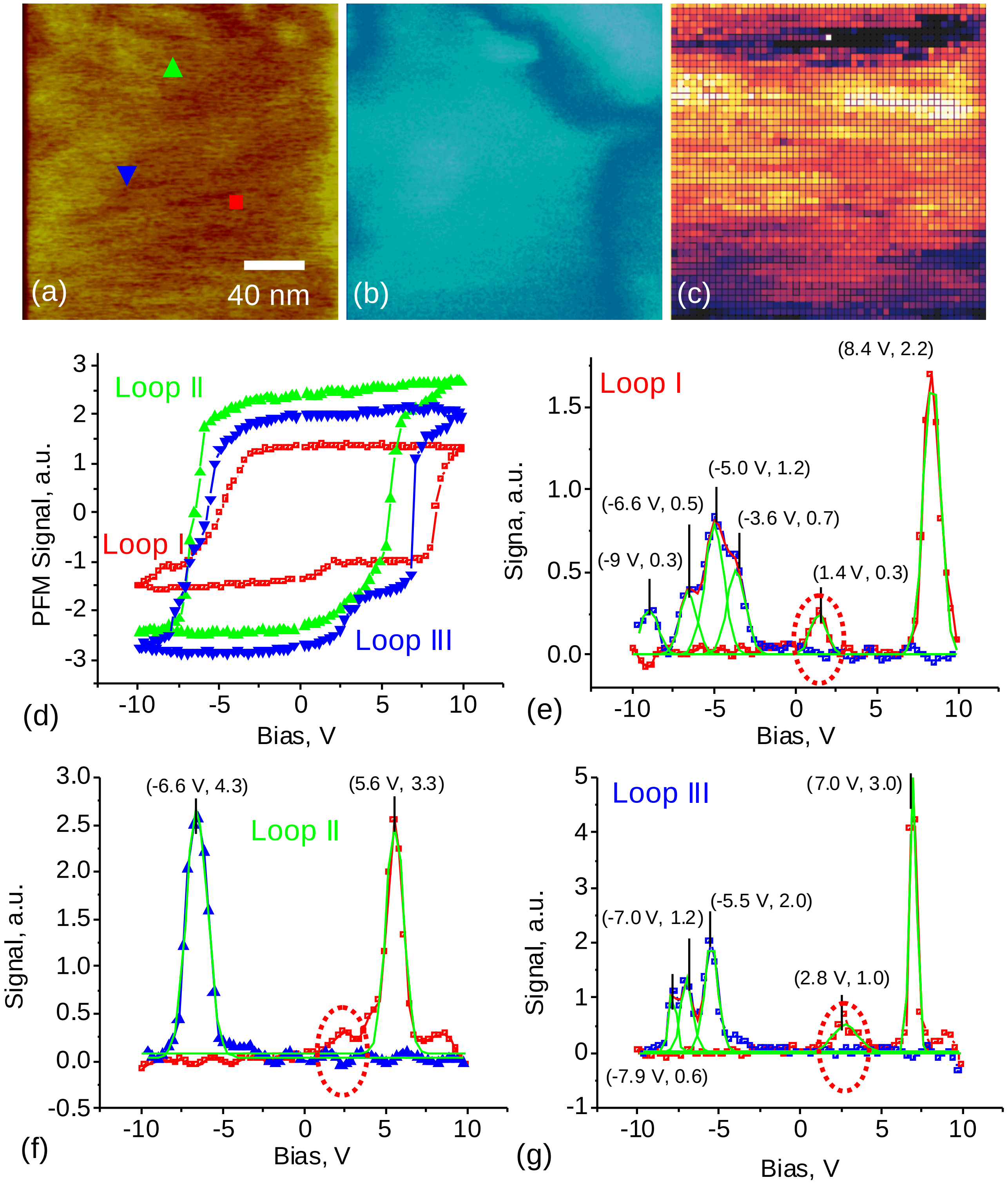

**Figure 1.** S.V. Kalinin, S. Jesse, B.J. Rodriguez et al.

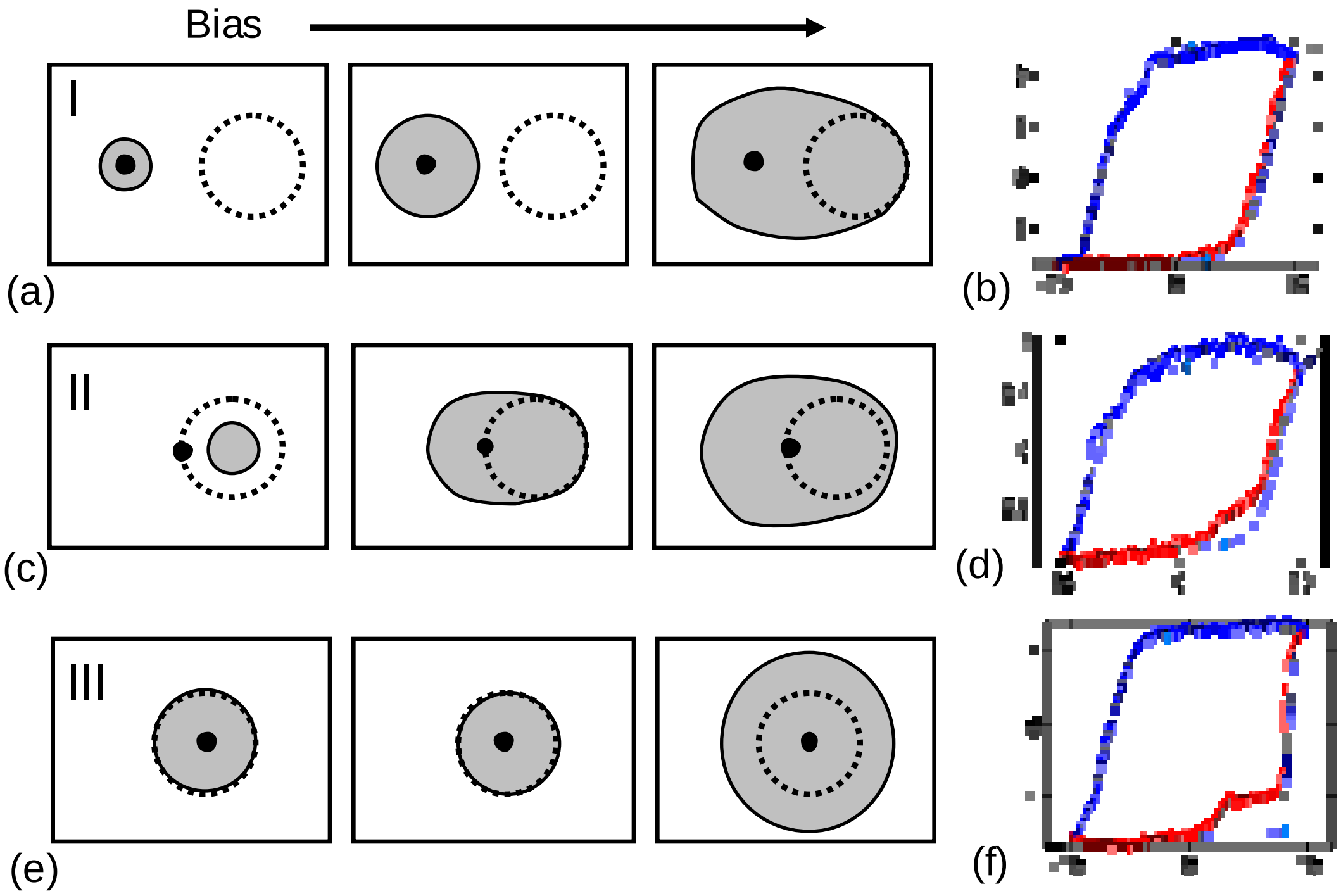

**Figure 2.** S.V. Kalinin, S. Jesse, B.J. Rodriguez et al.

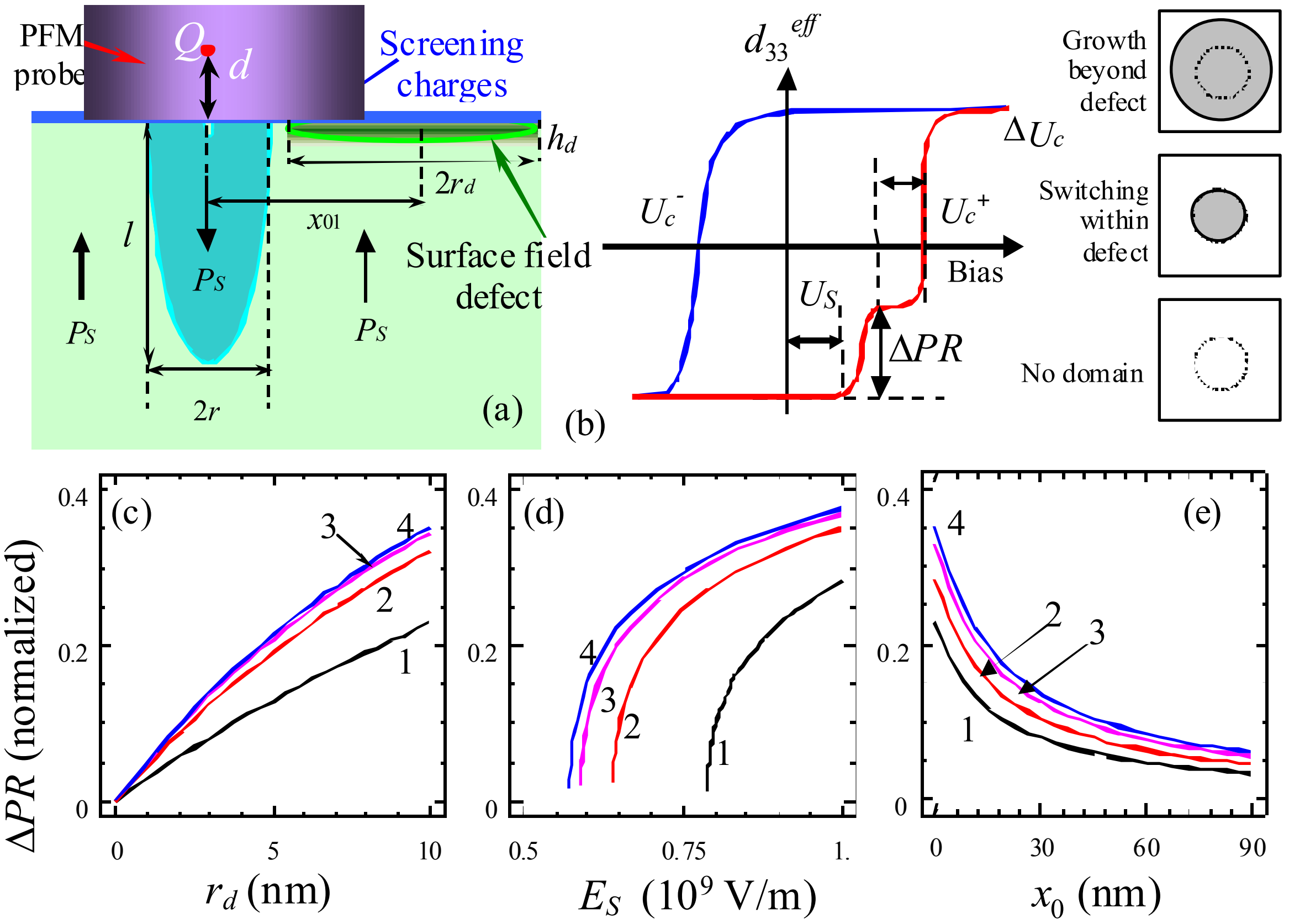

**Figure 3.** S.V. Kalinin, S. Jesse, B.J. Rodriguez et al.

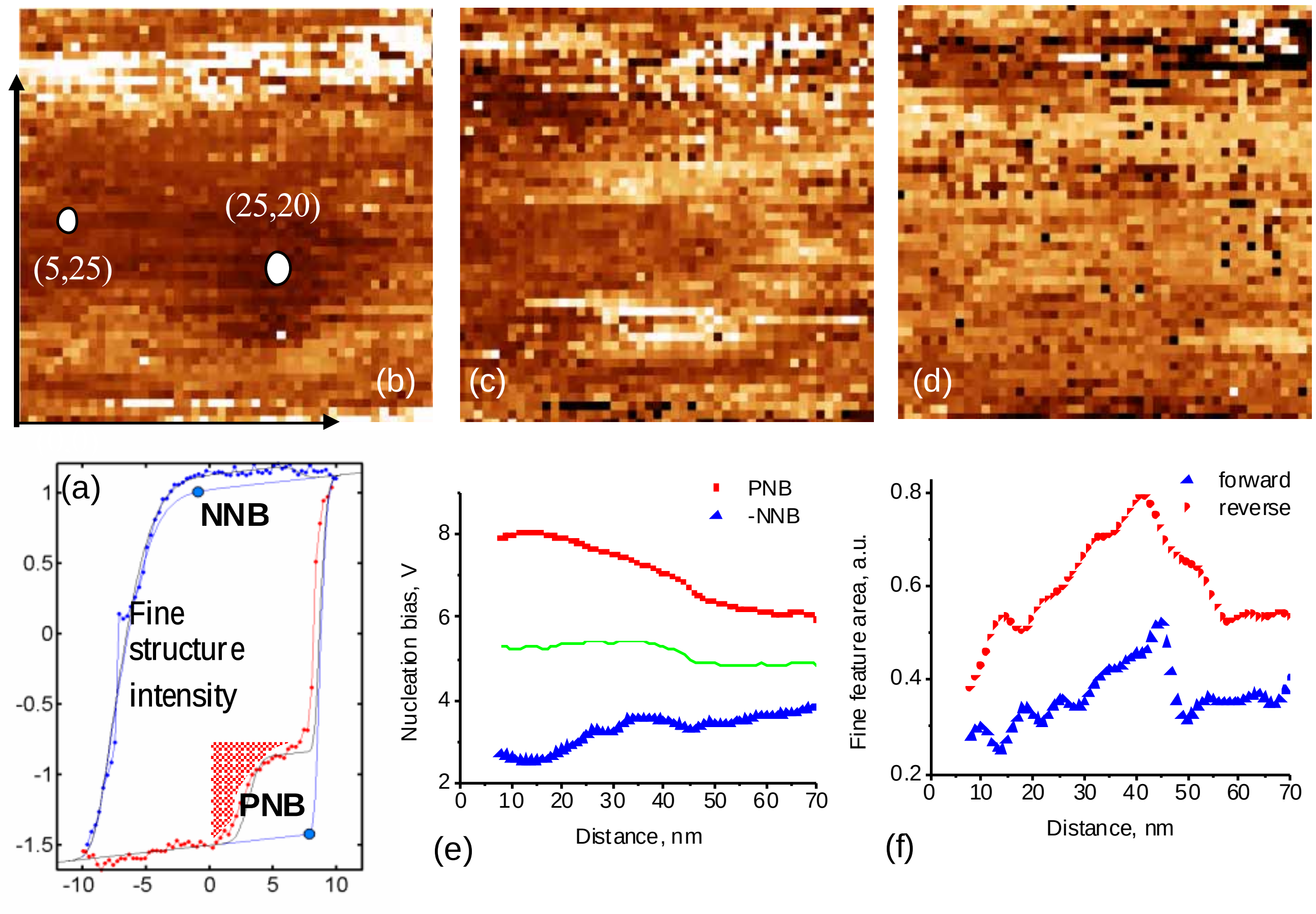

**Figure 4.** S.V. Kalinin, S. Jesse, B.J. Rodriguez et al.

*Supplementary materials for the manuscript.*

# Probing the role of single defects on thermodynamics of electric-field induced phase transitions

S.V. Kalinin,[1,2,†] S. Jesse,[1] and B.J. Rodriguez,[1,2]

[1]Materials Sciences and Technology Division and [2]The Center for Nanophase Materials Sciences, Oak Ridge National Laboratory, Oak Ridge, TN 37831

Y.H. Chu and R. Ramesh,

Department of Physics and Department of Materials Science and Engineering, University of California, Berkeley, CA 94720

E.A. Eliseev and A.N. Morozovska

National Academy of Science of Ukraine, Kiev, Ukraine

The analysis of the switching process in the presence of a surface dipolar defect is presented in Section I. The description of the data fitting procedures and definitions of measured parameters is given in Section II. Experimental SS-PFM data, representative hysteresis loops, and full sets of 2D parameters describing switching properties for the $BiFeO_3$ film studied in this work are presented in Section III.

† Corresponding author, sergei2@ornl.gov

## I. Effect of a charged defect on domain formation

The domain nucleation caused by the charged PFM probe is analyzed within the framework of Landauer-Molotskii model [28, 29]. The excess of surface and electrostatic energy of the semi-ellipsoidal domain is:

$$\Phi(r,l) \cong \psi_S S(r,l) - \frac{P_S}{2}\int_V dv\left(E_3^p(\mathbf{x}) + E_3^d(\mathbf{x})\right) + \frac{4\pi l\, r^2}{3\varepsilon_0\varepsilon_{11}} n_D(r,l) P_S^2 \ , \qquad (1)$$

where $r$ is the domain radius, $l$ is its length, $S$ –surface area, $\psi_S$ - is the domain wall energy density, $P_S$ is the magnitude of spontaneous polarization. The rigorous expression for the depolarization factor $n_D(r,l)$ is given in [30]. The electric field established by the probe is $E_3^p(\mathbf{x}) = -\nabla\varphi_p(\mathbf{x})$, and the electric field created by defects is $E_3^d(\mathbf{x}) = -\nabla\varphi_d(\mathbf{x})$ (random frozen disorder). The depolarization field energy is calculated under the condition of perfect tip-surface electric contact or/and surface screening by free charges.

For SS-PFM maps analyses, the relevant expansion basis for the resulting defect random electric field $E_3^d(\mathbf{x}) = \sum_i E_{di}(\mathbf{x})$ should be chosen. To develop the analytical description of defect-mediated switching, we adopt the Gaussian basis and suppose that any physical electric field is finite and thus could be expanded on the basis [31]:

$$E_3^d(\mathbf{x}) = \sum_i E_{Si} \exp\left( -\frac{(x - x_{0i})^2 + (y - y_{0i})^2}{r_{di}^2} - \frac{z}{h_{di}} \right) \qquad (2)$$

So, the $i$-th Gaussian basis element electric field has the amplitude $E_{Si}$, characteristic radius $r_{di}$, penetration depth $h_{di} << r_{di}$ and $\mathbf{x}_{0i} = \{x_{0i}, y_{0i}, 0\}$ is the center location with respect to the coordinate origin $\{0,0,0\}$ linked with the sample surface. Further analysis is performed assuming that the semi-ellipsoidal domain is axi-symmetric, but allowing for defects influence the domain center $\mathbf{b}=\{b_1,b_2,0\}$ (i.e. the variational vector describing domain position) is shifted on distance $|\mathbf{a}-\mathbf{b}|$ compared to the PFM tip apex location at a given scanning point $\mathbf{a}=\{a_1,a_2,0\}$.

So, the free energy of a semi-ellipsoidal domain is:

$$\Phi(r,l,\mathbf{b},U)\approx\left(\begin{array}{l}\pi\psi_S\,l\,r\left(\dfrac{r}{l}+\dfrac{\arcsin\sqrt{1-r^2/l^2}}{\sqrt{1-r^2/l^2}}\right)+\\ +\dfrac{P_S^2}{\varepsilon_0\varepsilon_{33}}\dfrac{4\pi r^2 l}{3}\dfrac{(r\gamma/l)^2}{1-(r\gamma/l)^2}\left(\dfrac{\mathrm{arctanh}\left(\sqrt{1-(r\gamma/l)^2}\right)}{\sqrt{1-(r\gamma/l)^2}}-1\right)-\\ -\dfrac{4\pi U P_S\,d\,r^2\,l/\gamma}{\left(\sqrt{r^2+d^2+|\mathbf{a}-\mathbf{b}|^2}+d\right)\left(\sqrt{r^2+d^2+|\mathbf{a}-\mathbf{b}|^2}+d+\dfrac{l}{\gamma}\right)}-\\ -2\pi P_S\sum_i r_{di}^2 h_{di}E_{Si}I_S\left(r,|\mathbf{x}_{0i}-\mathbf{b}|,r_{di},h_{di}\right)\end{array}\right)\tag{3a}$$

Here $U$ is the voltage applied to the probe, $d$ is the effective charge-surface separation (i.e. effective tip parameter). In the effective point charge approximation, the probe is represented by a single charge $Q=2\pi\varepsilon_0\varepsilon_e R_0U(\kappa+\varepsilon_e)/\kappa$ located at $d=\varepsilon_e R_0/\kappa$ for a spherical tip, or $d=2R_0/\pi$ for a flattened tip represented by a disk in contact. Parameters $\gamma=\sqrt{\varepsilon_{33}/\varepsilon_{11}}$ and $\kappa=\sqrt{\varepsilon_{33}\varepsilon_{11}}$ are the effective dielectric constant and anisotropy factor.

The dimensionless overlap integral $I_S(r,x,r_d,h_d)$ between the domain and given basis element has the form

$$\begin{aligned}I_S(r,x,r_d,h_d)&\approx 2\left(1-\exp\left(-\frac{l}{h_d}\right)\right)\int_0^{r/r_d}\rho d\rho\cdot I_0\left(2\frac{x}{r_d}\rho\right)\exp\left(-\rho^2-\frac{x^2}{r_d^2}\right)\approx\\ &\approx\left(1-\exp\left(-\frac{l}{h_d}\right)\right)\left(1-\exp\left(-\frac{r^2}{r_d^2}\right)\right)\exp\left(-\frac{x^2}{r_d^2}\right)\end{aligned}\tag{3b}$$

In order to obtain approximate analytical expressions for the tip-induced domain nucleation, let us simplify the free energy, Eqs. (3), for several special cases. Under the conditions $l >> d$, $l >> h_{di}$ and $r < d$, valid for initial stages of stable domain growth in the presence of large-scale defects with $r_{di} > r$, Eq. (3a) can be rewritten as:

$$\Phi \approx \left(\begin{array}{l} \pi\psi_S\, l r\left(\dfrac{r}{l}+\dfrac{\arcsin\sqrt{1-r^2/l^2}}{\sqrt{1-r^2/l^2}}\right)- \\ -2\pi r^2 P_S\left(\dfrac{2d\cdot U}{\sqrt{d^2+|\mathbf{a}-\mathbf{b}|^2}+d}+\sum_i h_{di}E_{Si}\exp\left(-\dfrac{|\mathbf{x}_{0i}-\mathbf{b}|^2}{r_{di}^2}\right)\right)+ \\ +\dfrac{P_S^2}{\varepsilon_0\varepsilon_{33}}\dfrac{4\pi r^2 l}{3}\dfrac{(r\gamma/l)^2}{1-(r\gamma/l)^2}\left(\dfrac{\operatorname{arctanh}\left(\sqrt{1-(r\gamma/l)^2}\right)}{\sqrt{1-(r\gamma/l)^2}}-1\right) \end{array}\right) \tag{4}$$

The term $\dfrac{\sqrt{d^2+|\mathbf{a}-\mathbf{b}|^2}+d}{d}\sum_i h_{di}E_{Si}\exp\left(-\dfrac{|\mathbf{x}_{0i}-\mathbf{b}|^2}{r_{di}^2}\right)$ acts as a built-in voltage and thus determines the left-to-right imprint $\Delta U_a = U_a^+ + U_a^-$ of domain piezoelectric response loop nucleation bias $U_a^\pm \cong \pm\dfrac{4}{3}\gamma d\left(\sqrt{\dfrac{2\pi\psi_S^3}{3E_a P_S^2}}+\dfrac{|P_S|}{3\varepsilon_0\varepsilon_{11}}\right)+\Delta U_a$ measured at the given scanning point $\mathbf{a}=\{a_1,a_2,0\}$, $E_a$ is the activation barrier.

Then the simplest analysis of the free energy (3) maps was performed for a single Gaussian defect with center at $\{x_{01},0,0\}$ and tip apex located in coordinate origin $\{0,0,0\}$. Corresponding dependence of the activation barrier, domain radius, length and piezoresponse on applied voltage for the different distances to the single Gaussian defect is shown in Fig.S1 for $BiFeO_3$. It is clear that the activation barrier $E_a$ is almost absent for bias $U \ge 0$ in the vicinity of a positive surface field defect $|x_{01}|/r_d < 1$ with the field strength $E_S \cong 10^8 - 10^{10}$ V/m, so that surface domain state exists [e.g. see Fig. S1]. Opposite to the tip-induced formation of stable domains, spontaneous surface domain state is pinned by the nearest defect site, i.e. $\mathbf{b} \approx \{x_{01}, 0, 0\}$. After numerical minimization of Eq. (3) with respect to **b** we obtained, that the tip-induced domain center shift towards the defect exponentially vanishes for $|x_{01}|/r_d > 1$, as anticipated. For positive bias $U \geq 0$ and negative surface field defect ( $|x_{01}|/r_d < 1$ and e.g. $E_S = -(10^8 - 10^{10})$ V/m) no surface state exists and the activation barrier drastically increases, as should be expected.

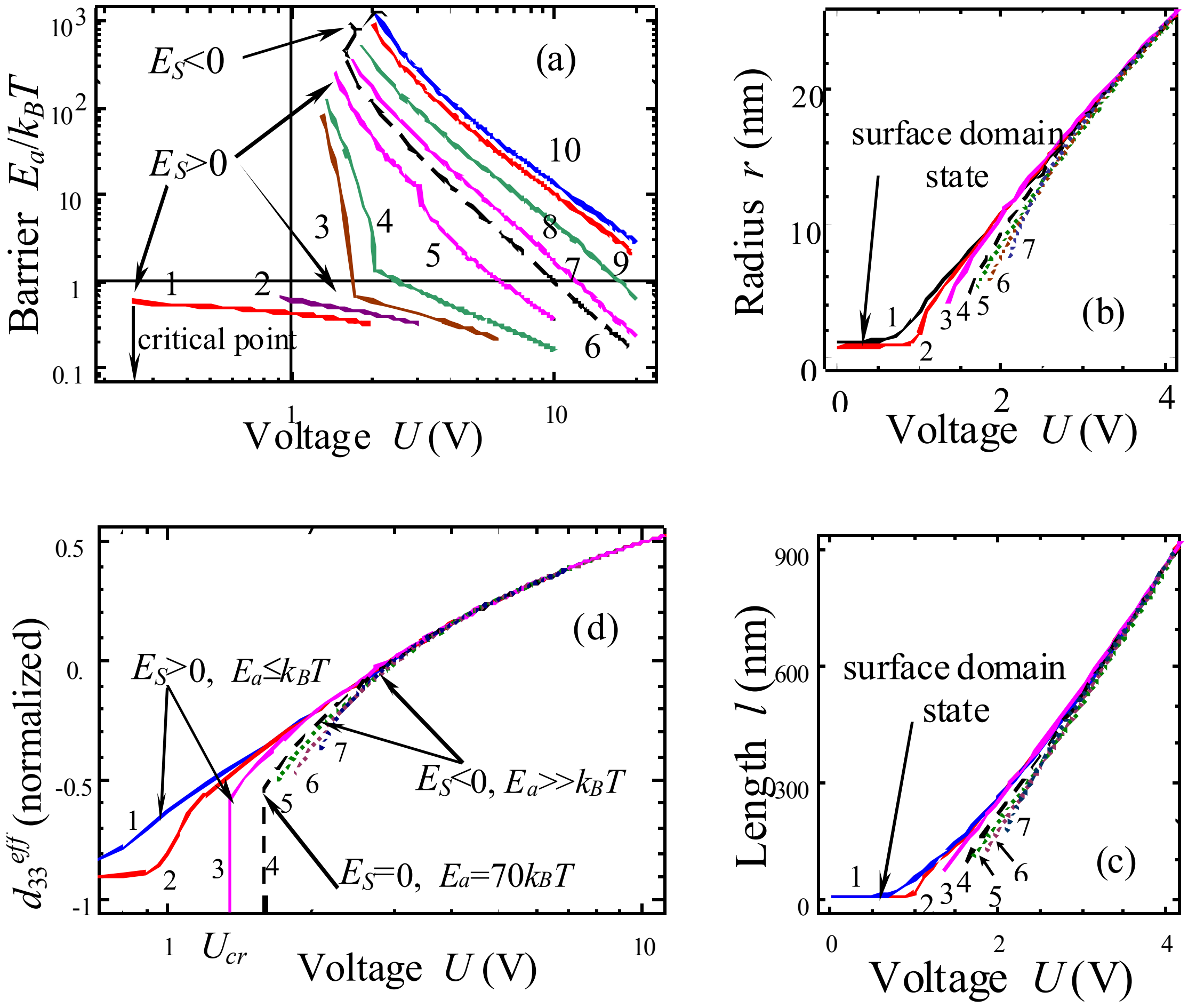

**Fig. S1.** (a) Dependence of activation energy $E_a$ (in $k_BT$ units) on the applied voltage $U$ for the different distance to the single Gaussian defect $x_{01}$ = 8, 8.5, 9, 10, 13 nm for $E_S = 5\cdot 10^8$ V/m (curves 1 - 5), $E_S = 0$ (curve 6) and $x_{01}$ = 15, 10, 5, 0 nm for $E_S = -5\cdot 10^8$ V/m (curves 7 - 10). Dependence of domain radius (b) and length (c) on the applied voltage for the different distances $x_{01} = 0$, 5, 10nm, and $E_S = 5\cdot 10^8$ V/m (curves 1 - 3); $E_S = 0$ (curve 4) and 15, 10, 0 nm for $E_S = -5\cdot 10^8$ V/m (curves 5 - 7). (d) Dependence of normalized PFM response loop on the applied voltage for the different distance to the defect $x_{01}$ = 0, 5 nm, $\infty$, and $E_S = 5\cdot 10^8$ V/m (solid, long- and short-dashed curves)

calculated for nucleation bias $U_a^{\pm}$ at level $E_a = (0.5-1)$ eV and week pinning limit. Defect characteristic radius $r_d = 10$ nm, penetration depth $h_d = 0.8$ nm. Material parameters used hereinafter correspond to $BiFeO_3$ (polarization $P_S$ = 0.5 C/m$^2$, permittivity $\varepsilon_{33} \approx 70$ , dielectric anisotropy $\gamma \approx 1$, domain wall surface energy $\psi_S$ = 20 mJ/m$^2$; $d_{33}$ = 25pm/V, $d_{31}$ = -10pm/V, $d_{15}$ = 20pm/V; point charge-surface separation $d$ = 30 nm in total charge approximation for disk-plane model).

For a given defect distribution, numerical calculations assure us, that at typical activation barriers $E_a$=2-20$k_B T$ the tip-induced formation of stable domains appears below the tip apex, i.e. $|\mathbf{a}-\mathbf{b}| << d$ and so the left-to-right coercive bias imprint $\Delta U_C = U_C^{+} + U_C^{-}$ is

$$\Delta U_C(a_1, a_2) \approx -2\sum_i h_{di} E_{Si} \exp\left( -\frac{(x_{0i}-a_1)^2 + (y_{0i}-a_2)^2}{r_{di}^2} \right). \qquad (5)$$

From Eq. (5), the coercive bias imprint $\Delta U_C$ exponentially decreases with distance from defect center. Using single defect parameters $h_d$ =(2-0.5) nm, $r_d \geq$(5-10) nm and typical field strength $|E_S| \cong 10^8 - 10^{10}$ V/m [32], we estimate that $|\Delta U_C| = (0.2-5)$ V. For dielectric anisotropy $\gamma \approx 1$, corresponding equilibrium sizes of prolate domains are:

$$r(U) \approx \frac{6\varepsilon_0\varepsilon_{11}}{25\pi\psi_S}\left( \ln\left( \frac{8U P_S}{5\pi\psi_S} + \frac{4\Delta U_C P_S}{5\pi\psi_S} \right) - 1 \right)^{-1} \left( U + \frac{\Delta U_C}{2} \right)^2, \qquad (6a)$$

$$l(U) \approx \frac{24\varepsilon_0\varepsilon_{11} P_S}{125\pi^2 \psi_S^2}\left( \ln\left( \frac{8U P_S}{5\pi\psi_S} + \frac{4\Delta U_C P_S}{5\pi\psi_S} \right) - 1 \right)^{-1} \left( U + \frac{\Delta U_C}{2} \right)^3, \qquad (6b)$$

In particular case of equal $h_{di} \approx h_d$ that corresponds to the same screening conditions of all surface field defects, in accordance with Eq.(2), Eq.(5) is exactly proportional the surface field in the scanning point $\mathbf{a} = \{a_1, a_2, 0\}$, namely:

$$E_3^d(a_1, a_2, z=0) = \sum_i E_{Si} \exp\left(-\frac{(a_1 - x_{0i})^2 + (a_2 - y_{0i})^2}{r_{di}^2}\right) = -\frac{\Delta U_C(a_1, a_2)}{2h_d} \qquad (7)$$

Performed numerical simulations proved, that under the conditions $l >> d$ , $l >> h_{di}$ and $r >> d$ , valid for saturation stage of domain growth, and aforementioned random field strength the domain sizes become insensitive to the presence of defects.

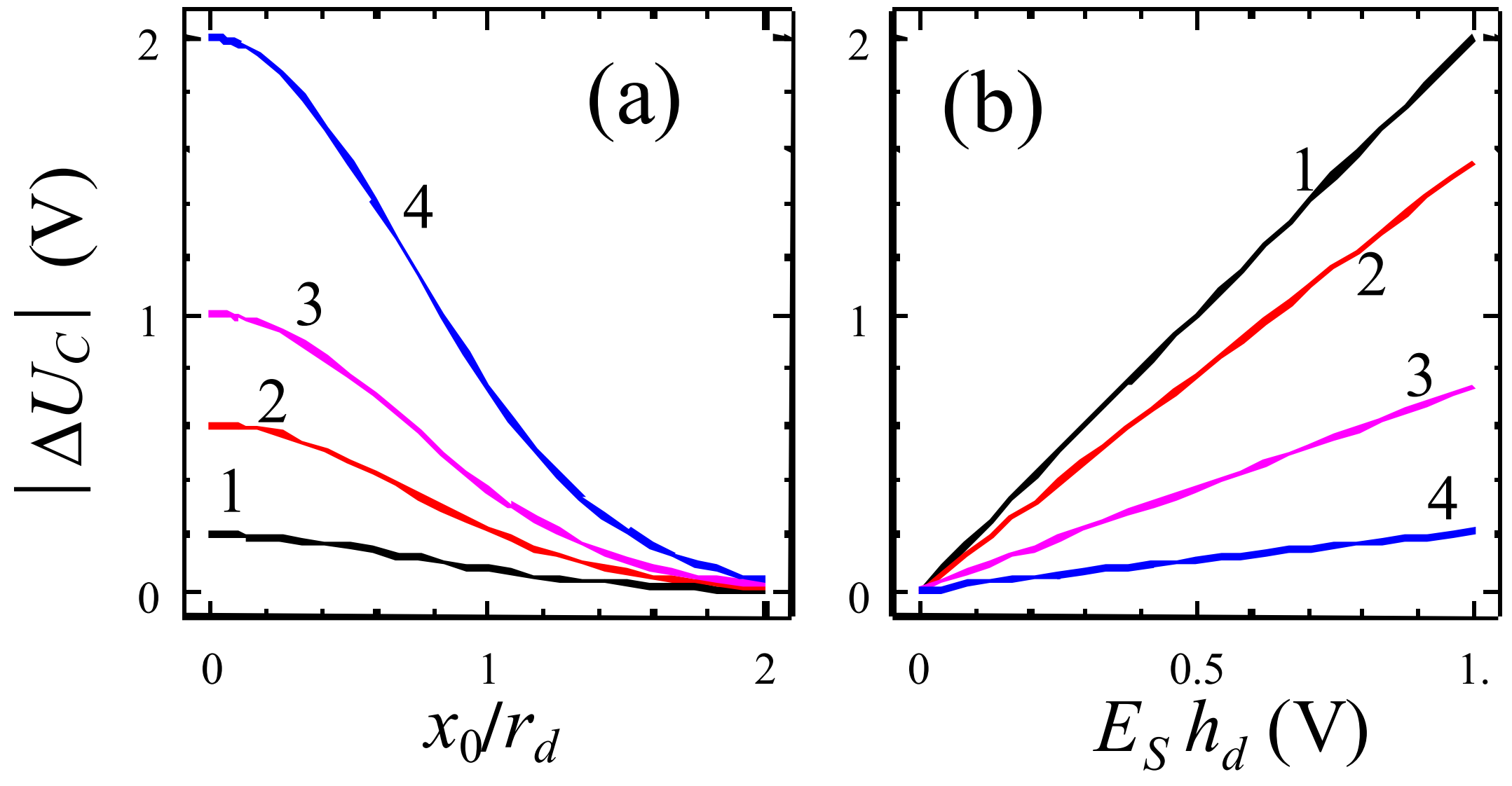

**Fig. S2.** The role of single Gaussian defect parameters $\{E_S, h_d, r_d, x_0\}$ on piezoresponse loop asymmetry $|\Delta U_C|$: (a) $|E_S|$ $h_d$=0.1, 0.3, 0.5, 1 V (curves 1, 2, 3, 4); (b) $x_0/r_d$=0, 0.5, 1, 1.5 (curves 1, 2, 3, 4). Material parameters correspond to $BiFeO_3$ and listed in Fig. S1.

Further analysis of the free energy (3) has shown that under the presence of several well-separated strong field defects an oblate sub-surface domain with polarization sign $P_S E_{Sm} > 0$, sizes $l \sim h_{dm} << d$ and $r >> l$ is pinned by the nearest defect site $\mathbf{x}_{0m} = \{x_{0m}, y_{0m}, 0\}$. Under the conditions $l \le h_{dm} << r_{dm}$, $l << d$, valid for surface domain state, Eqs. (3) lead to:

$$\Phi \approx \left( \begin{array}{l} \pi\psi_S\, l\, r\left(\frac{r}{l} + \frac{\arcsin\sqrt{1-r^2/l^2}}{\sqrt{1-r^2/l^2}}\right) + \frac{4\pi r^2 l}{3\varepsilon_0\varepsilon_{33}}\frac{P_S^2 (r\gamma/l)^2}{1-(r\gamma/l)^2}\left(\frac{\operatorname{arctanh}\left(\sqrt{1-(r\gamma/l)^2}\right)}{\sqrt{1-(r\gamma/l)^2}} - 1\right) \\ \frac{-4\pi P_S r^2 l\cdot d\cdot U}{\gamma\left(\sqrt{d^2+|\mathbf{a}-\mathbf{b}|^2}+d\right)^2} - 2\pi P_S E_{Sm} l r_{dm}^2\left(1-\exp\left(-\frac{r^2}{r_{dm}^2}\right)\right)\exp\left(-\frac{(\mathbf{x}_{0m}-\mathbf{b})^2}{r_{dm}^2}\right) \end{array} \right) \qquad (8)$$

Under the condition $\mathbf{b} = \mathbf{x}_{0m}$, valid for strongly pinned surface domain, its radius can be estimated as:

$$r_{Sm}(U) \approx r_{dm}\sqrt{\ln(2h_{dm}P_S E_{Sm}) - \ln\left(\psi_S + \frac{4h_{dm}P_S^2}{3\varepsilon_0\varepsilon_{33}} - \frac{4P_S h_{dm}\, d\cdot U}{\gamma\left(\sqrt{d^2+(\mathbf{x}_{0m}-\mathbf{a})^2}+d\right)^2}\right)} \qquad (9)$$

For the spontaneous surface state appearance at $U$=0 the inequality $P_S E_{Sm} > \frac{\psi_S}{2h_{dm}} + \frac{2P_S^2}{3\varepsilon_0\varepsilon_{33}}$ should be valid. As is follows from Eq.(9), the surface domain state becomes absolutely unstable at bias

$$U_S^{cr} = \frac{\gamma}{4P_S d}\left(\sqrt{d^2+(\mathbf{x}_{0m}-\mathbf{a})^2}+d\right)^2\left(\frac{\psi_S}{h_{dm}} + \frac{4P_S^2}{3\varepsilon_0\varepsilon_{33}} - 2E_{Sm}P_S\right) \qquad (10)$$

At higher absolute values of applied bias only the tip-induced domain formation can be stable. However, overestimated bias (10) is not the characteristic bias of fine structure peculiarities appearance/disappearance, since the sub-surface domain center location $\mathbf{b}$ does not coincide with defect site $\mathbf{x}_{0m} \neq \mathbf{a}$ under applied bias increase. More correct, but still rather rough estimation for the characteristic bias is $U_S \approx \frac{\gamma}{4P_S d}\left(\sqrt{d^2+|\mathbf{a}-\mathbf{b}|^2}+d\right)^2\left(\frac{\psi_S}{h_{dm}}+\frac{4P_S^2}{3\varepsilon_0\varepsilon_{33}}-2P_S E_{Sm}\exp\left(-\frac{|\mathbf{x}_{0m}-\mathbf{b}|^2}{r_{dm}^2}\right)\right)$, where the position $\mathbf{b}$ should be determined from the free energy minimum.

The surface domain state is responsible for the corresponding step-like piezoresponse loop fine structure (e.g. the one depicted in Fig.1-4). When the tip position is just at the defect site (i.e. $|\mathbf{x}_{0m}-\mathbf{a}| << d$) the loop fine structure is one-sided and appears at the bias $U_S^+$, while the other side nucleation is activationless (i.e. $U_a^- \approx 0$). Since the PFM response in the center of a cylindrical domain is $d_{33}^{eff}(r) \approx \frac{3}{4}d_{33}^*\frac{\pi d-8r}{\pi d+8r}+\frac{d_{15}}{4}\frac{3\pi d-8r}{3\pi d+8r}$, where $d_{33}^* = d_{33}+\frac{d_{31}}{3}(1+4\nu)$ ($\nu$ is the Poisson ratio) [33], the ratio $r_{Sm}/d$ determines the piezoresponse fine structure vertical jump $\Delta d_{33}^{eff}(r_{Sm}) = d_{33}^{eff}(0)-d_{33}^{eff}(r_{Sm})$ (see e.g. the step depicted in Fig.3d). Under the condition $|a_1 - x_{0m}| << d$ and $|a_2 - y_{0m}| << d$ the relative jump of the piezoelectric response is:

$$\Delta PR_S(r_{Sm}, \mathbf{a}=\mathbf{x}_{0m}) = \frac{d_{33}^{eff}(0)-d_{33}^{eff}(r_{Sm})}{d_{33}^{eff}(0)} \approx \frac{16 r_{Sm}}{3d_{33}^*+d_{15}}\left(\frac{3d_{33}^*}{\pi d+8r_{Sm}}+\frac{d_{15}}{3\pi d+8r_{Sm}}\right), \qquad (11a)$$

The jump $\Delta PR(r_{Sm})$ decreases with increasing distance between the tip apex and defect center $|\mathbf{a}-\mathbf{x}_{0m}|$ allowing for electric field vanishing, at that Pade approximation is valid:

$$\Delta PR_S\left(r_{Sm},|\mathbf{a}-\mathbf{x}_{0m}|\right)\approx\frac{16r_{Sm}}{3d_{33}^{*}+d_{15}}\left(\frac{3d_{33}^{*}}{\pi d+8\left(r_{Sm}+|\mathbf{a}-\mathbf{x}_{0m}|\right)}+\frac{d_{15}}{3\pi d+8\left(r_{Sm}+|\mathbf{a}-\mathbf{x}_{0m}|\right)}\right), \quad (11b)$$

Using Eqs. (5-11), it is potentially possible to reconstruct defects parameters and distribution from the piezoresponse loops coercive bias imprint $\{\Delta U_{Cn}\}$, fine structure maximal jumps $\{\Delta PR_{Sn}^{m}\}$, characteristic bias of fine structure peculiarities $\{U_{Sn}^{m}\}$ experimentally measured in the number $n=1...N$ of scanning points $\mathbf{a}^n=\{a_1^n,a_2^n,0\}$, the number $m=1...M$ indicates several possible fine structure elements for a given $n$-th loop.

**II. Fitting procedure and automated data analysis**

Data analysis is performed using a phenomenological fitting function possessing characteristic elements of an ideal hysteresis loop with additional parameters that can be activated to account for fine structure variations at the nucleation corners of the loop. The least squared error fitting process is performed in a likewise two step process in which the data is fitted first to the ideal, 9 parameter, description for the purposes of gross estimation, after which fine features (accounting for an additional 4 parameters) are considered. This two tier process is advantageous in both speed and accuracy in that the likelihood for the fitting algorithm confusing gross with fine structure is greatly reduced.

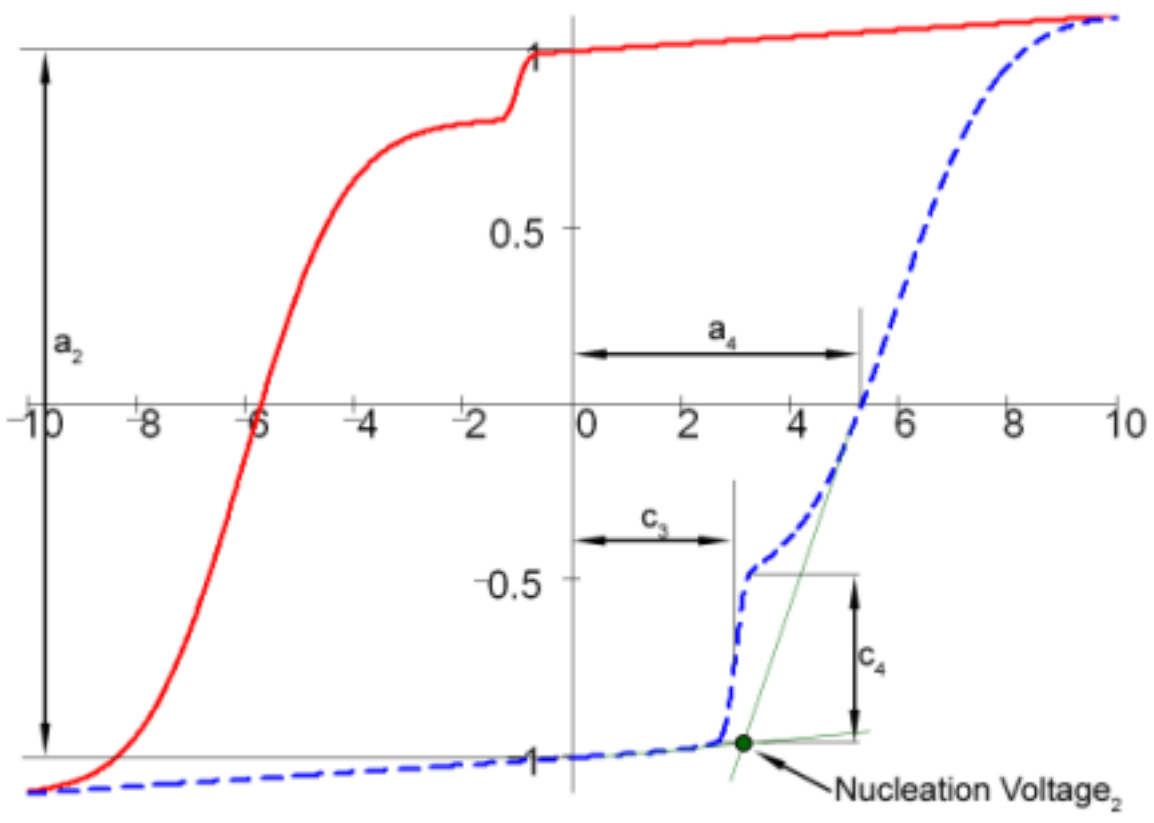

**Fig. S3**. Representative 13 parameter fitting function with fitting parameters $a_{1\text{-}5}$ = [ -1 2 -6 6 .01], $b_{1\text{-}4}$ = [ 2 2 2 2], and $c_{1\text{-}4}$ = [ -1 .1 3 .3 ].

Each branch of the 9 parameter fitting function, □$_1$ and □$_2$, is defined as an error function with a continuously varying standard deviation, □*(x)*, which transitions between two values across the center of the error function. The fitting variables $b_{1\text{-}4}$ (always >0) define the sharpness the corners of the ideal hysteresis loop. □ is much smaller than the coercive bias range. $a_{1\text{-}4}$ define the horizontal and vertical positions of the centers of each branch and $a_5$ is the slope of the saturated regions of the curves.

Fine structures are accounted for with smaller error functions, □*(x)*, located near each corner of the ideal hysteresis loop. The centers of the fine structure steps are located at $c_1$ and $c_3$. The height of each step (measured as a fraction of the total height of the entire hysteresis loop) are determined by $c_2$ and $c_4$.

$$\sigma_1(x) = \frac{(b_2 - b_1)}{2} \cdot \left( erf\left( \frac{x - a_3}{\alpha} \right) + 1 \right) + b_1 \qquad \text{(I.1)}$$

$$\sigma_2(x) = \frac{(b_4 - b_3)}{2} \cdot \left( erf\left( \frac{x - a_4}{\alpha} \right) + 1 \right) + b_3 \qquad \text{(I.2)}$$

$$\varsigma_1(x) = \frac{c_2}{2} \cdot \left( erf\left( \frac{x - c_1}{\beta} \right) + 1 \right) \quad \text{(I.3)}$$

$$\varsigma_2(x) = \frac{c_4}{2} \cdot \left( erf\left( \frac{x - c_3}{\beta} \right) + 1 \right) \quad \text{(I.4)}$$

$$\Gamma_1(x) = a_1 + \frac{a_2}{1 + c_2} \cdot \left[ \frac{\sigma_1(x)}{(b_1 + b_2)} \cdot erf\left( \frac{x}{\sigma_1(x)} \right) + \frac{b_1}{(b_1 + b_2)} + \varsigma_1(x) \right] + a_5 \cdot x \quad \text{(I.5)}$$

$$\Gamma_2(x) = a_1 + \frac{a_2}{1 + c_4} \cdot \left[ \frac{\sigma_2(x)}{(b_3 + b_4)} \cdot erf\left( \frac{x}{\sigma_2(x)} \right) + \frac{b_3}{(b_3 + b_4)} + \varsigma_2(x) \right] + a_5 \cdot x \quad \text{(I.6)}$$

The relationship between the phenomenological fitting parameters and characteristic parameters of the hysteresis loop is following: $V^- = a_4$, $V^+ = a_3$, imprint is $I = (a_3 - a_4)/2$, $R^- = a_1 - a_2$, $R^+ = a_1$, $slope = a_5$, switchable polarization is $SP = a_2$, work of switching is $Work = |a_3 - a_4| \cdot |a_2|$, positive nucleation bias $V_{n1} = a_4 + 2b_4$, positive nucleation bias $V_{n2} = a_3 - 2b_1$. Area of the fine structures are estimated as $W_{fs1} = |V_{n1} - a_3| \cdot c_2$ and $W_{fs2} = |V_{n2} - a_4| \cdot c_4$. An example fit is shown in figure S2 for the fitting parameters $a_{1\text{-}5}$ = [ -1 2 -6 6 .01], $b_{1\text{-}4}$ = [ 2 2 2 2], and $c_{1\text{-}4}$ = [ -1 .1 3 .3 ].

## III. Experimental data for $BiFeO_3$ film

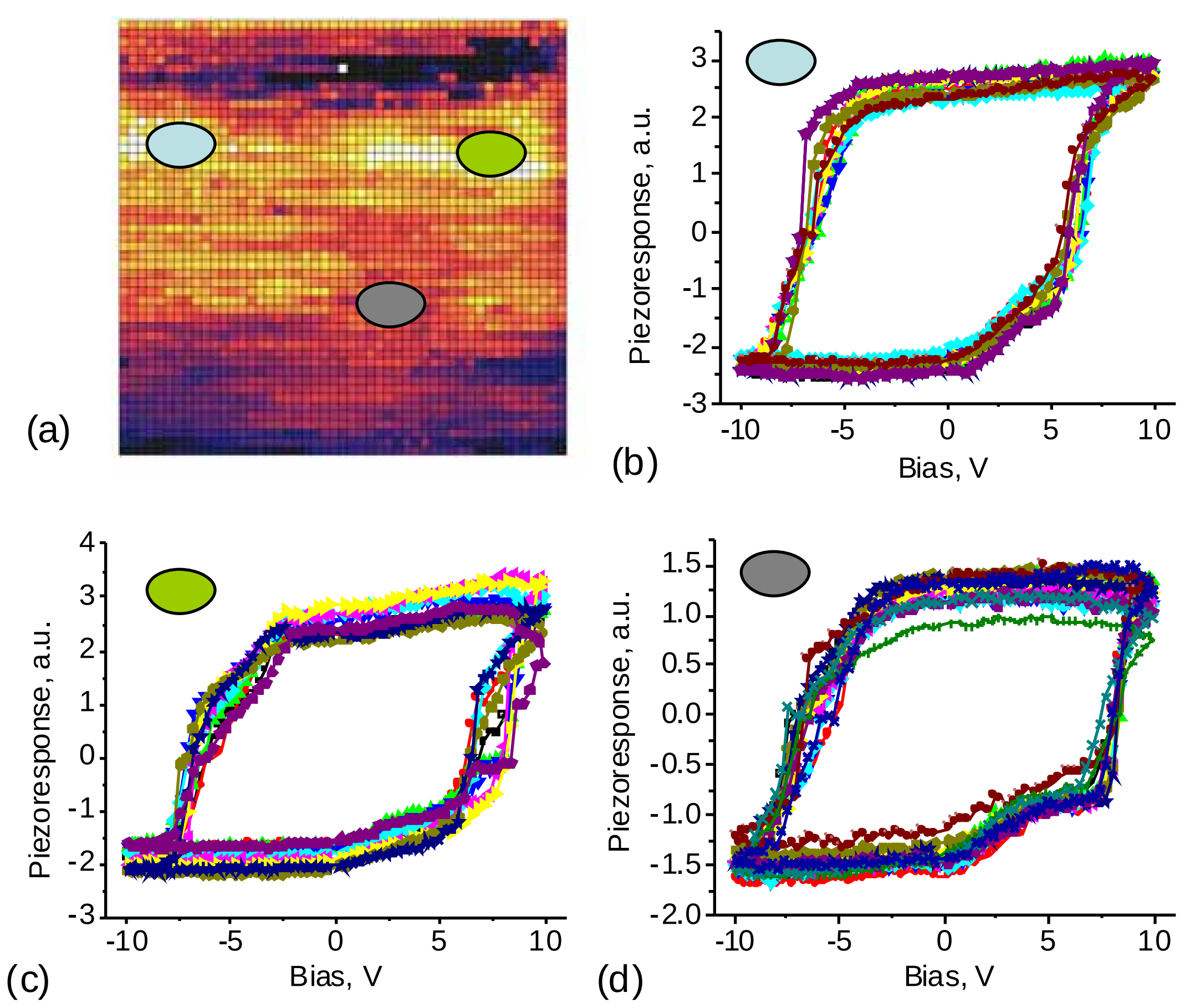

**Fig. S4.** (a) SS-PFM initial response map. (b) – (d) show multiple hysteresis loops measured across the regions indicated in (a). Note the consistency of the fine structure within these regions and strong variations between the regions.

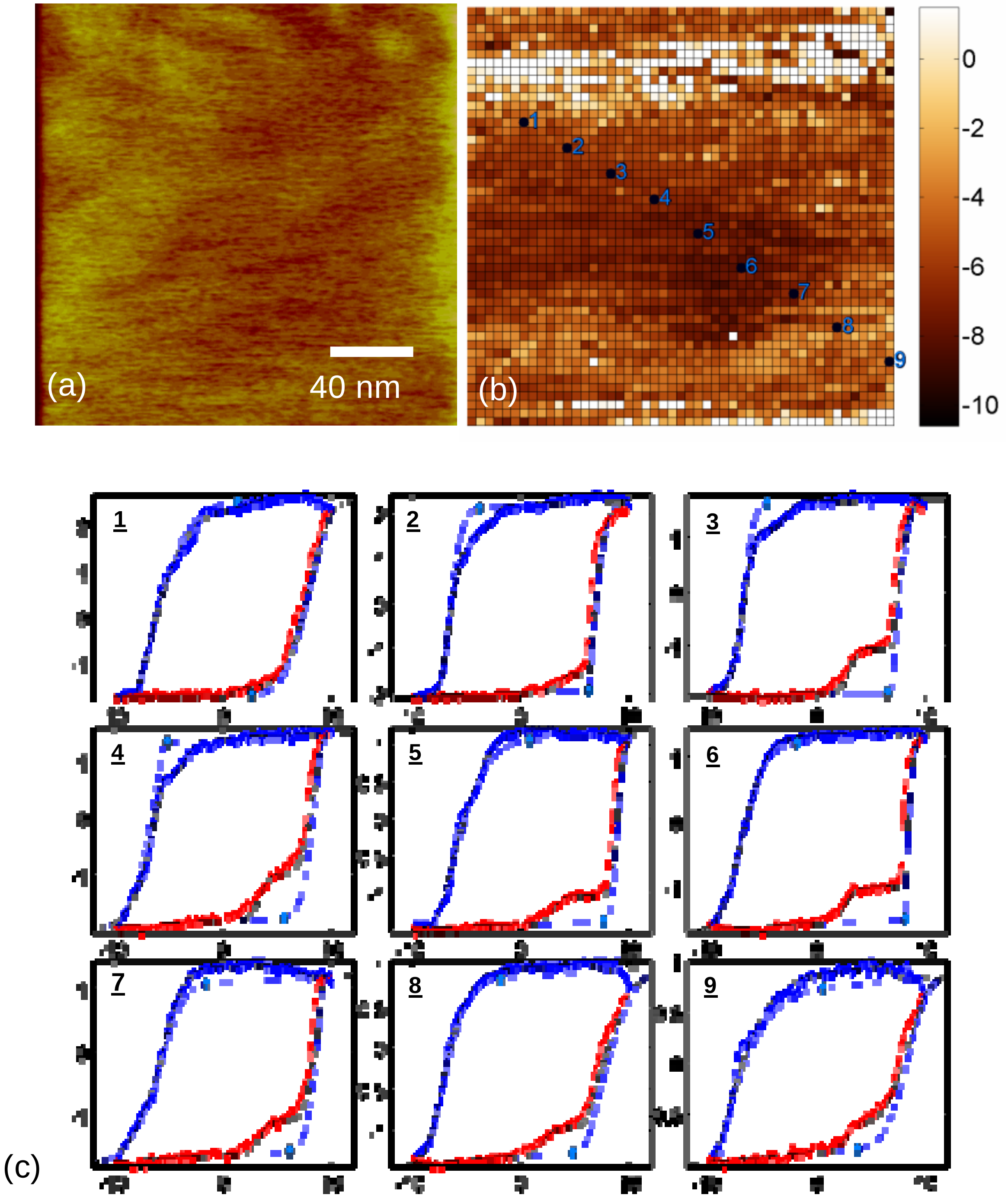

**Fig. S5.** (a) Topography and (b) nucleation bias in the vicinity of a defect. The hystereisis loops acquired at various points in (b) are shown in (c). Note the gradual variation in hysteresis loop fine structure while crossing the defect. The 13 parameter fit is shown in black,

the 9 parameter (ideal) fit is shown in blue with the ideal nucleation biases indicated by blue dots.

**Figure S6.** Maps of ferroelectric switching parameters describing large scale hysteresis structure.

**Figure S7.** Maps of fine structure switching parameters. $c_1$ and $c_3$ indicate the biases at which the fine structures are first detected. $c_2$ and $c_4$ are the heights (measured as a fraction of the total hysteresis loop height) of the fine structures.

SS-PFM initial response image and reproducibility in the hysteresis loop shape between the loops collected from adjacent locations is shown in Fig. S4. The full set of the 2D SS-PFM maps describing overall loops structure is shown in Fig. S5. The SS-PFM maps for the fine structure are shown in Fig. S6,7.

[23] E.M. Lifshitz, L.D. Landau, and L.P. Pitaevskii, *Electrodynamics of Continuous Media*, Butterworth Heinemann, 1984.

[24] G. Gerra, A.K. Tagantsev, N. Setter, Phys. Rev. Lett. **94,** 107602 (2005)

[25] A.N. Morozovska, S.V. Svechnikov, E.A. Eliseev, B.J. Rodriguez, S. Jesse, and S.V. Kalinin, submitted, also http://arxiv.org/abs/0711.1426

[26] M. Rief, F. Oesterhelt, B. Heymann B and H.E. Gaub, Science **275**, 1295 (1997).

[27] C. Bustamante, Y.R. Chemla, N.R. Forde, D. Izhaky, Annu. Rev. Biochem. **73**, 705 (2004).

[28] . R. Landauer, Electrostatic Considerations in $BaTiO_3$ Domain Formation during Polarization Reversal. J. Appl. Phys. **28**, 227 (1957).

[29] M. Molotskii, J. Appl. Phys. **93**, 6234 (2003).

[30] E.M. Lifshitz, L.D. Landau, and L.P. Pitaevskii, *Electrodynamics of Continuous Media*, Butterworth Heinemann, 1984.

[31] The expansion is invalid the point of e.g. dipole defect location, but the dipole approximation itself is valid far from the charges itself.

[32] G. Gerra, A.K. Tagantsev, N. Setter, PRL **94,** 107602 (2005)

[33] A.N. Morozovska, E. Eliseev, S. Bravina, and S.V. Kalinin, submitted